\documentclass[a4paper,11pt]{article}
\usepackage{jcappub} 

\title{\boldmath Map-based E/B separation of filtered timestreams using space-based E-mode observations}

\author[a]{Yuyang Zhou}
\author[a,b]{Adrian Lee}
\author[c,d]{Yuji Chinone}

\affiliation[a]{Department of Physics, University of California, Berkeley, Berkeley, CA 94720, USA}
\affiliation[b]{Physics Division, Lawrence Berkeley National Laboratory, Berkeley, CA 94720, USA}
\affiliation[c]{International Center for Quantum-field Measurement Systems for Studies of the Universe and Particles (QUP), High Energy Accelerator Research Organization (KEK), Tsukuba, Ibaraki 305-0801, Japan}
\affiliation[d]{Kavli Institute for the Physics and Mathematics of the Universe (WPI), UTIAS, The University of Tokyo, Kashiwa, Chiba 277-8583, Japan}

\emailAdd{yuyang\_zhou@berkeley.edu}
\emailAdd{Adrian.Lee@berkeley.edu}
\emailAdd{chinoney@gmail.com}

\abstract{E to B mixing or “leakage" due to time-ordered data (TOD) filtering has become an important source of sensitivity loss that ground-based cosmic microwave background polarization experiments must address. However, it is a difficult problem for which very few viable solutions exist. In this paper, we expand upon satellite E-mode methods to cover E/B leakage specifically due to TOD filtering. We take a satellite E-mode map and TOD filter it through the ground-based experiment data analysis pipeline, from which we construct a map-space ``leakage template" and subtract it from the ground-based experiment map. We evaluate the residual leakage by simulating the satellite E-mode maps with Planck-like and LiteBIRD-like noise levels, and simulate the ground-based experiment with Simons Observatory-like and CMB-S4-like noise levels. The effectiveness of the method is measured in the improvement of the Fisher uncertainty $\sigma(r=0)$. We find that our method can reduce $\sigma(r=0)$ by $\sim15\text{--}75\%$ depending on the noise levels considered.}

\begin{document}
\maketitle
\flushbottom

\section{Introduction} \label{sec:intro}
The cosmic microwave background (CMB) is our most powerful observational probe of physical conditions in the early universe. Although its spectrum and temperature anisotropies have been exquisitely measured by numerous experiments \cite{cobe, wmap, planck}, its polarization anisotropies remain a frontier waiting to be fully characterized. In particular, the B-mode component of the polarization (defined in Section~\ref{subsec:eb decomposition}) is expected to contain the imprint of the stochastic background of gravitational waves predicted by theories of cosmic inflation \cite{Seljak_1997}. The yet undetected amplitude of this signal depends sensitively on the energy scale of inflation, providing a powerful constraint on the space of inflationary models. A precise measurement of the primordial B-mode polarization thus lies at the forefront of modern cosmology.  

As experiments become more sensitive in this pursuit, systematics previously buried underneath the noise spectrum start to become relevant. One such effect is the leaked E-mode variance into the B-mode spectrum due to time-domain filtering of the polarization time-ordered data (TOD), so called E-to-B (E/B) leakage. This occurs in ground-based experiments that use filter-bin mapmaking pipelines \cite{Hivon_2002}, which is expected to be the method of choice for large datasets \cite{s4}. Since E-modes are much brighter than the primordial B-mode signal, any leaked variance will dramatically affect the overall BB sensitivity. It is worth noting that E/B leakage due to TOD filtering is not present when using both maximum likelihood mapmaking and maximum likelihood power spectrum estimation. However, this is only computationally feasible at low resolutions \cite{Hivon_2002}.

\begin{figure}[htbp] 
\centering
\includegraphics[width=0.6\linewidth]{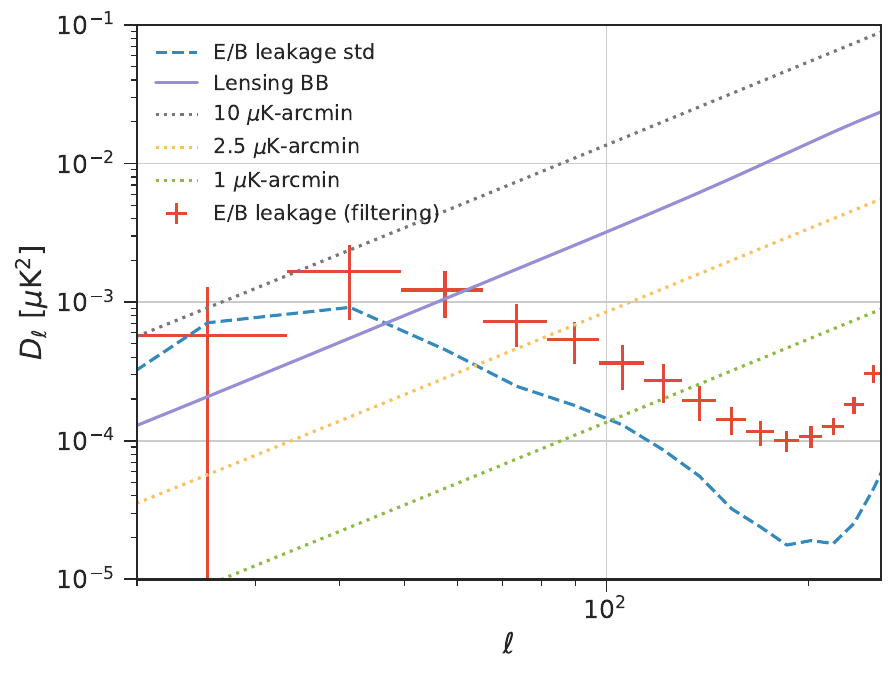}
\caption{Comparison between the BB leakage spectra (red) due to TOD filtering and a few different white noise levels. The leakage spectra is estimated from filtering 128 EE maps with a standard subscan polynomial and ground template subtraction. The mean is plotted in each bin with standard deviation error bars with the standard deviation also plotted separately in the blue dashed line. We see that the leakage spectrum, concentrated at low-$\ell$ due to the nature of the filtering, becomes dominant over the noise spectra for more sensitive experiments.} 
\label{fig:leakage}
\end{figure}

In the past, ground-based experiments have handled E/B leakage due to TOD filtering using Monte Carlo (MC) simulations. A suite of signal-only E-mode maps is generated from $\Lambda$CDM input and TOD filtered through the analysis pipeline. The mean BB pseudospectra of those filtered E-mode maps is subtracted from the data BB pseudospectrum (recent example in \cite{Adachi_2020}). While this successfully debiases the data BB mean, it does not remove the variance of the much brighter E-modes, resulting in decreased BB sensitivity at low multipoles. For experiments achieving a map depth below a few $\mu$K-arcmin, this effect will be dominant over the noise spectrum (Figure~\ref{fig:leakage}) and thus will need to be treated in a more sophisticated way.  

The only successful demonstration of mitigating E/B leakage due to TOD filtering on a real dataset is that of ``matrix purification," a method implemented by the BICEP/Keck collaboration \cite{Ade_2016} at the South Pole. However, while effective, this method is computationally demanding and can be challenging to implement for experiments with increasing detector count and Chilean scan strategies \cite{scan} (see discussion in Section~\ref{subsec:matpure}). Nevertheless, the matrix purification approach is an avenue currently being explored by the Simons Observatory (SO) \cite{so} based in Chile.

In this paper, we propose a simpler, less computationally demanding alternative by extending the work of \cite{Ghosh_2021}. We leverage a single, deep E-mode map with sufficient low-$\ell$ coverage (such as a space mission) and TOD filter it through the ground-based experiment data analysis pipeline to estimate a ``leakage template" in map space. While the method presented in \cite{Ghosh_2021} mitigates E/B leakage due to partial sky coverage (Section~\ref{subsec:partial sky coverage}), our method extends this to cover E/B leakage due to TOD filtering (Section~\ref{subsec:TOD filtering}) for which there are currently no easily implementable solutions. To evaluate noise scaling, we simulate Planck \cite{planck} and LiteBIRD \cite{litebird} noise levels for the satellite E-mode maps and SO and CMB-S4 (S4) \cite{s4} noise levels for the ground-based experiment maps. Finally, we forecast the improvement in $\sigma(r)$ and compare the results to the existing matrix purification method. 

\section{E/B Mixing} \label{sec:eb mixing}
\subsection{E/B Decomposition} \label{subsec:eb decomposition}
Experiments measure CMB polarization in terms of the Stokes parameters Q/U, which are related to the underlying polarization field via the spin-2 spherical harmonics \cite{Bunn_2003} 
\begin{equation}
    Q \pm iU = \sum_{l,m} {}_{\pm2}a_{\ell m} \, _{\pm2}Y_{l,m}. 
\end{equation}
Since Stokes Q/U depend on the coordinate system, power spectrum results are usually presented in a rotationally invariant basis, denoted E and B. The corresponding coefficients are defined as  
\begin{equation}
\begin{aligned}
    a_{\ell m}^E &\equiv -(_{2}a_{\ell m} + _{-2}a_{\ell m}) / 2 \\
    a_{\ell m}^B &\equiv -i(_{2}a_{\ell m} - _{-2}a_{\ell m}) / 2
\end{aligned}
\end{equation}
and similarly the spherical harmonics (in vector notation) 
\begin{equation}
\begin{aligned}
    Y_{\ell m}^E &\equiv 
    \biggl( 
        \begin{aligned}
            (_{2}Y_{\ell m} + _{-2}Y_{\ell m}) / 2 \\ 
            -i(_{2}Y_{\ell m} - _{-2}Y_{\ell m}) / 2
        \end{aligned}
    \biggl) \\
    Y_{\ell m}^B &\equiv 
    \biggl( 
        \begin{aligned}
            i(_{2}Y_{\ell m} - _{-2}Y_{\ell m}) / 2 \\
            (_{2}Y_{\ell m} + _{-2}Y_{\ell m}) / 2 
        \end{aligned}
    \biggl)
\end{aligned}
\end{equation}
with 
\begin{equation}
    \int_{S^2} Y_{\ell m}^E \cdot Y_{\ell' m'}^B \, dS = 0
\end{equation}
over the full sphere \cite{Ade_2016}. The overall polarization field can then be written as 
\begin{equation}
    \mathbf{P} = 
    \biggl(
        \begin{aligned}
            Q \\ 
            U 
        \end{aligned}
    \biggl)
    = -\sum_{l,m} a_{\ell m}^E \, Y_{\ell m}^E + a_{\ell m}^B \, Y_{\ell m}^B.
\end{equation}
where it is straightforward to see that E and B-only polarization fields are orthogonal over the full sky
\begin{equation} \label{equ:ortho}
\int_{S^2} \mathbf{P}^E \cdot \mathbf{P}^B \, dS = 0.
\end{equation}
That is, there is no E/B mixing when we have a complete measurement of the sky signal. 

When we do not have a complete measurement (i.e. information loss), we have E/B mixing. The two main sources relevant for ground-based experiments are partial-sky coverage and TOD filtering, in order of impact.

\subsection{Partial Sky Coverage} \label{subsec:partial sky coverage}
Ground-based observatories naturally only have access to part of the sky, causing Equation~\ref{equ:ortho} to be no longer integrated over the full sphere. In this case, the spaces of E and B-modes begin to mix, with the overlap comprising ``ambiguous" modes, which contribute power to both EE and BB.

Techniques for dealing with E/B mixing due to partial sky coverage have been extensively studied, with the most widely used being \cite{Smith_2006, grain}, which we will refer to as ``KS purification" after the author. For the simulation results in the rest of this paper, we make use of the NaMaster pseudo-$C_\ell$ package's \cite{Alonso_2019} implementation of KS purification (see Appendix~\ref{sec:ks_impl}). 

\subsection{TOD Filtering} \label{subsec:TOD filtering}
All ground-based experiments employ some form of TOD filtering to suppress low-frequency noise, such as the atmosphere, ground emission, and thermal drifts. Due to the scanning direction of the telescope, the filtering runs along a specific direction on the sky. This means that different $Y_{\ell m}$ modes are affected differently by filtering operations depending on their orientation, breaking the orthogonality relation in Equation~\ref{equ:ortho} and inducing E/B mixing.          
For Stage-2 experiments such as POLARBEAR \cite{Adachi_2020} targeting $\sigma(r)\sim10^{-2}$, the E/B leakage caused by TOD filtering was subdominant to the noise. Thus, only the mean level of leakage was estimated from filtered EE simulations and subtracted from the pseudospectra via
\begin{equation} \label{equ:leakage}
    \Tilde{C}_{\ell}^{E \rightarrow B} = \frac{F_{\ell}^{E \rightarrow B}}{F_{\ell}^{E \rightarrow E}} \Tilde{C}_{\ell}^{E}
\end{equation}
where $F_{\ell}$ is the filter transfer function approximated as the ratio of output over input power spectra. Although this correctly debiases the central value of BB in each bin, the leaked variance of the E-modes is left in the data.

For Stage-3/4 experiments such as SO \cite{so} and S4 \cite{s4} targeting $\sigma(r)\sim10^{-3}$ and below, the leaked E-mode variance is now dominant over the noise spectrum, prompting the development of new techniques to deal with this now relevant effect. The canonical solution has become the BICEP/Keck collaboration's matrix purification method \cite{Ade_2016}, which we briefly review in the next section.

\subsection{Matrix Purification} \label{subsec:matpure}
We present an overview of the matrix purification approach and defer additional details to \cite{Ade_2016}. The method involves finding ``pure-B" modes that remain orthogonal to the E-modes after pipeline processing (masking, filtering, etc.). We seek to construct a new basis that satisfies
\begin{equation}
    \mathbf{R}\mathbf{P}^E \cdot \mathbf{b} = 0
\end{equation}
where \textbf{R} is a linear operator called the ``observation matrix" and \textbf{b} is the new pure-B basis. \textbf{R} represents the application of the entire data analysis pipeline, including data selection, filtering, apodization, etc. We will henceforth refer to maps processed through the pipeline onto a sky patch as ``observed" maps. $\mathbf{b}$ is constructed by solving the generalized eigenvalue problem (omitting the regularization factor for clutter, see Appendix \ref{sec:matpure impl})
\begin{equation} \label{equ:gep}
    \Tilde{\mathbf{C}}_B \mathbf{x}_i = \lambda_i \Tilde{\mathbf{C}}_E \mathbf{x}_i
\end{equation}
where $\Tilde{\mathbf{C}}_{E/B}$ are the ``observed" signal covariances
\begin{equation} \label{equ:obs_cov}
    \Tilde{\mathbf{C}}_{E/B} = \mathbf{R}^\intercal 
        \left(
            \sum_{l,m} C_{\ell}^{EE/BB} Y^{E/B}_ {\ell m} Y^{E/B \dagger}_{\ell m}
        \right)
    \mathbf{R}.
\end{equation}
The unobserved signal covariances in the brackets above are generated analytically following Appendix A of \cite{Tegmark_2001}. The solved eigenvectors of Equation~\ref{equ:gep} form our new basis of pure-B modes $\mathbf{b}$, with the eigenvectors with larger eigenvalues being more ``pure" than the eigenvectors with smaller eigenvalues. To see this, consider multiplying Equation \ref{equ:gep} from the left by $\mathbf{x_i^\intercal}$. A large eigenvalue $\lambda_i$ implies 
\begin{equation} \label{equ:eig_mag}
    \mathbf{x_i}^\intercal\Tilde{\mathbf{C}}_B \mathbf{x_i} \gg \mathbf{x_i}^\intercal\Tilde{\mathbf{C}}_E \mathbf{x_i}.
\end{equation}
which is the expected orthogonality behavior of a pure-B mode $\mathbf{b}$. Lastly, a ``purification" matrix is constructed via a projection operator using the pure-B basis
\begin{equation} \label{equ:pure}
    \boldsymbol{\Pi}_B = \sum_i \mathbf{b}_i (\mathbf{b}_i^\intercal \mathbf{b}_i)^{-1} \mathbf{b}_i^\intercal.
\end{equation}
The application of $\boldsymbol{\Pi}_B$ to an observed map projects the map onto the space of pure-B modes  
\begin{equation} \label{equ:purify}
    \mathbf{m}_\mathrm{purified}= \boldsymbol{\Pi}_B \mathbf{m}_\mathrm{obs}
\end{equation}
thereby removing the B contribution of ambiguous modes and mitigating E/B leakage.

There are primarily two computational challenges associated with the matrix purification method. The first is the generation of the observation matrix, for which the computation time depends on a number of factors. A bare-bones simulation for the SO small aperture telescope (SAT)'s south patch running at a reduced resolution of $N_\mathrm{side}=128$ in the HEALPix\footnote{\url{https://healpix.sourceforge.io/}} scheme, for 1/8 of the detectors in the 145 GHz band, with scanning taken from 10 operational days, using a simple subscan polyfilter and ground template subtraction, takes about 1 hour on 4 NERSC Perlmutter CPU nodes\footnote{\url{https://docs.nersc.gov/systems/perlmutter/architecture/}}. Although this may not sound prohibitive, there are several complicating factors. First, the computation time scales linearly with the number of detectors and days of scanning. Second, it is important to preserve the sparsity of the observation matrix due to hardware memory limits. The minimal simulation above only uses a subscan polyfilter and ground template subtraction, resulting in a matrix density of about 5\%. Increasing to $N_\mathrm{side}=256$ at the same sparsity exceeds the 512 GB memory limit of a NERSC Perlmutter CPU node. On top of that, including additional filtering that target many detectors/long timescales (such as a common mode filter) radically destroys the sparsity, limiting the choice of the filter stack for more realistic scenarios. In addition to computation time, the large memory footprint of the observation matrix makes subsequent matrix operations cumbersome. 

The second challenge is solving the generalized eigenvalue problem in Equation~\ref{equ:gep}. For the implementation used in this paper at $N_\mathrm{side}=128$, \texttt{scipy.sparse.eigsh}\footnote{\url{https://docs.scipy.org/doc/scipy/reference/generated/scipy.sparse.linalg.eigsh.html}} takes around 30 hours to calculate 7250 eigenvectors on a local CPU cluster running Intel(R) Xeon(R) E5-2650 v4 with 48 logical cores. Despite each observed covariance matrix in Equation~\ref{equ:obs_cov} being less than 3 GB, the routine required more than 50 GB of memory to complete, hinting at exponentially larger memory allocations required for higher resolutions/density.    

Lastly, any changes made to the pipeline (data selection, filtering, etc.) means the process needs to be repeated all over again, making iteration computationally expensive. Given these complications, the feasibility of implementing matrix purification for SO is an ongoing area of study. The details of the implementation used in this paper are described in Appendix~\ref{sec:matpure impl}.  

\subsection{Map-based E/B Separation} \label{subsec:map eb separation}
Given the computational challenges of the matrix purification method, we now introduce an alternative map-based E/B separation approach based on \cite{Ghosh_2021}. We describe the basic procedure in this section and leave the details of the implementation for Section~\ref{sec:simulations}. 

The main idea is as follows: given a perfect, noiseless map of the E-modes, E/B mixing would essentially be solved. We could process the map through any arbitrary data analysis pipeline, get the resulting B-mode leakage map, and remove it from our data. In contrast to the method described in Equation~\ref{equ:leakage}, this removes the leakage in map space as opposed to power spectrum space, which retains the full anisotropic information of the data. 

\begin{figure}[htbp]
\centering
\includegraphics[width=0.6\linewidth]{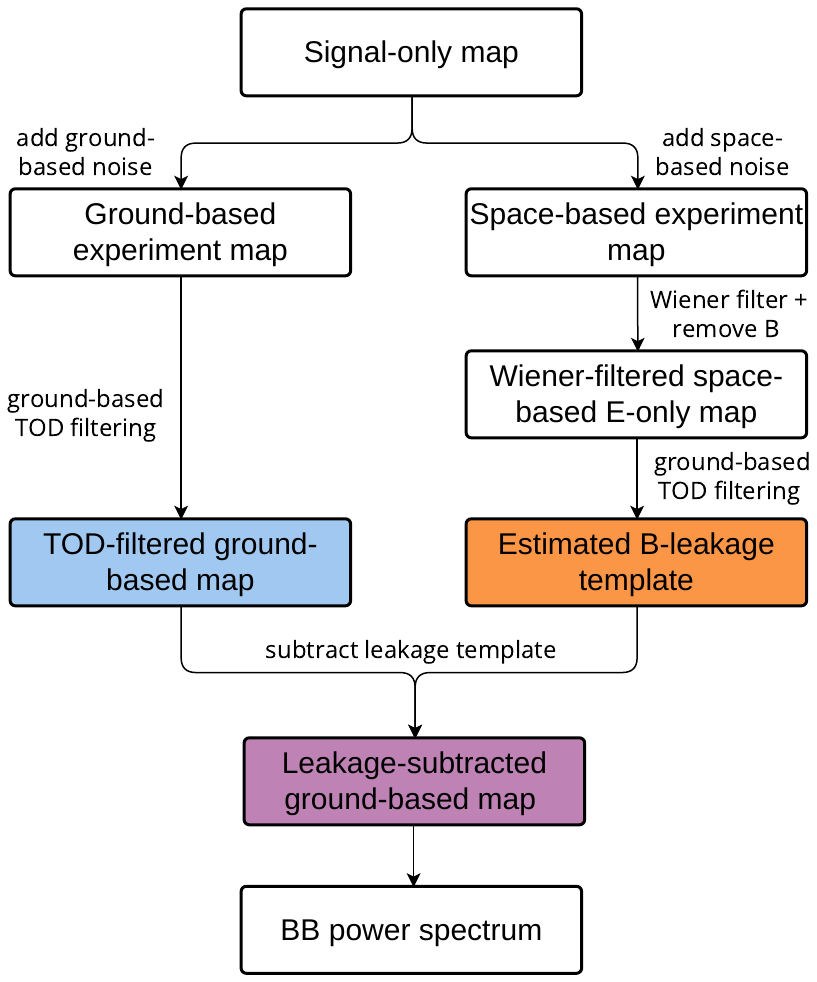}
\caption{Flowchart illustrating the process of map-based E/B separation. From a signal-only input map, we construct a ground-based experiment map with ground-based experiment noise and an E-mode only map with satellite experiment noise. The E-mode map is filtered through the same pipeline as the ground-based experiment, and the resulting B-mode map is used as a template for leakage removal. This process is repeated for each MC simulation.}
\label{fig:chart}
\end{figure}

Although a perfect E-mode map may never exist, we can nevertheless try to approximate this method by using existing noisy E-mode maps with sufficient low-$\ell$ coverage. One good candidate available now would be the Planck satellite \cite{planck} E-modes, or in the future the LiteBIRD satellite \cite{litebird} E-modes. With increasing noise levels of the satellite E-modes, our ability to estimate the E/B leakage template diminishes. The simulations in this paper aim to study the relationship between the satellite E-mode noise level and its effect on removing the E/B leakage incurred by TOD filtering. We simulate Planck-like and LiteBIRD-like noise levels for the satellite E-mode map and see if the residual leakage still dominates the sensitivity loss compared to the ground-based experiment noise level. See Figure~\ref{fig:chart} for a flowchart of the procedure.

Conceptually, the method is simple, which is a strength. Computationally, it would only require one additional run of the data analysis pipeline, which is trivial. It can also be done at any $N_\mathrm{side}$ without incurring cascading computational resources, unlike the matrix purification method. In the next section, we describe the simulation framework in more detail. 

\section{Simulations} \label{sec:simulations}
The simulation pipeline used to generate the results is made up of four parts corresponding to Figure~\ref{fig:chart}:
\begin{enumerate}
    \item $\Lambda$CDM signal-only maps 
    \item Ground-based experiment analysis pipeline + noise simulations 
    \item Satellite E-mode maps with isotropic noise 
    \item Power spectrum estimator
\end{enumerate}

To generate 1. $\Lambda$CDM signal-only maps, we first obtain a theory power spectra using CAMB with default Planck 2018 parameters\footnote{\url{https://github.com/cmbant/CAMB/blob/master/inifiles/planck_2018.ini}}.The simulated maps are then generated using \texttt{healpy.sphtfunc.synfast}\footnote{\url{https://healpy.readthedocs.io/en/latest/generated/healpy.sphtfunc.synfast.html}} with the theory power spectra as input. For the sake of simplicity, all maps are binned at $N_\mathrm{side}=128$ with smoothing based on pixel size, which ends up being roughly equivalent to a beam FWHM ($\theta_\mathrm{fwhm}$) of $\approx1^\circ$. Since we are mainly interested in studying the filtering leakage that manifests at low-$\ell$, this resolution is sufficient. In practice, maps from different experiments will need to be smoothed to a common beam before being combined. All simulations labeled ``EE/BB/EEBB" use $\Lambda$CDM input unless otherwise specified.

For 4. Power spectrum estimator, we use the NaMaster package \cite{Alonso_2019} to debias our pseudospectra and apply KS purification (for partial-sky leakage). The maps are apodized on the scale of $10^\circ$ in the ``C2" scheme. All debiased spectra shown are binned at $\Delta\ell$=16.

Items 2 and 3 will be discussed in more detail below.

\subsection{Ground-based Experiment Analysis Pipeline and Noise Simulations}
We construct a simple filter-bin pipeline to simulate our ground-based experiment maps, using the Time Ordered Astrophysics Scalable Tools (TOAST) ver. 3 software package \cite{theodore_kisner_2021_5559597}. All steps below are easily supported using built-in TOAST pipeline tools.

\begin{figure}[htbp]
\centering
\includegraphics[width=0.5\linewidth]{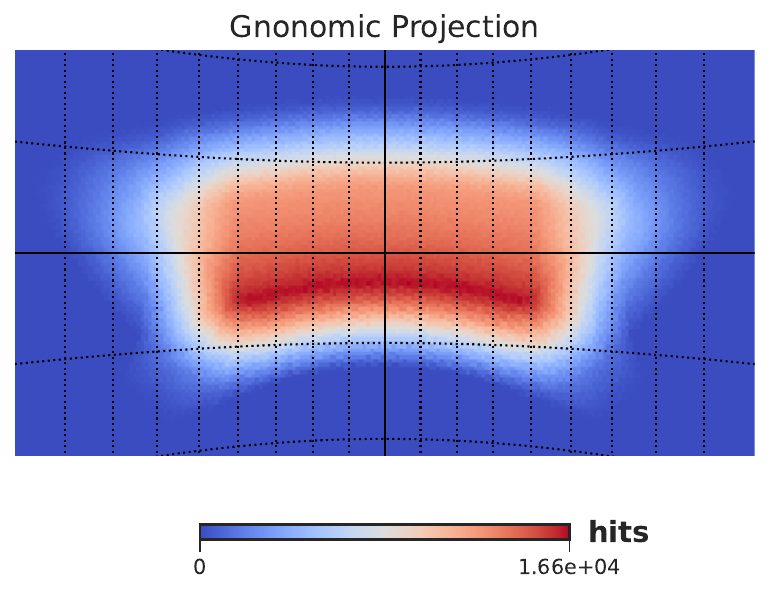}
\caption{Simulated ground-based experiment hit map over a 24 hour period. Covers a sky fraction of about 4.4\% in the southern field.}
\label{fig:hits}
\end{figure}

First, we generate a short 24-hour observing schedule at the Chilean Cerro Toco site targeting part of the southern field between 15 $<$ RA $<$ 65 and -50 $<$ DEC $<$ -30. This covers about 4.4\% of the sky, compared to the S4 deep patch (3\%) \cite{s4} and the SO SAT patch (10\%) \cite{so}. The observing schedule scans the patch twice a day (rising and setting) with constant elevation scans at different azimuths, which is a classic scan strategy from Chile (also used by SO \cite{so_scan}). Figure~\ref{fig:hits} shows the hit map for a simulated focal plane with a small number of detectors.

\begin{figure}[htbp]
\centering
\includegraphics[width=0.9\linewidth]{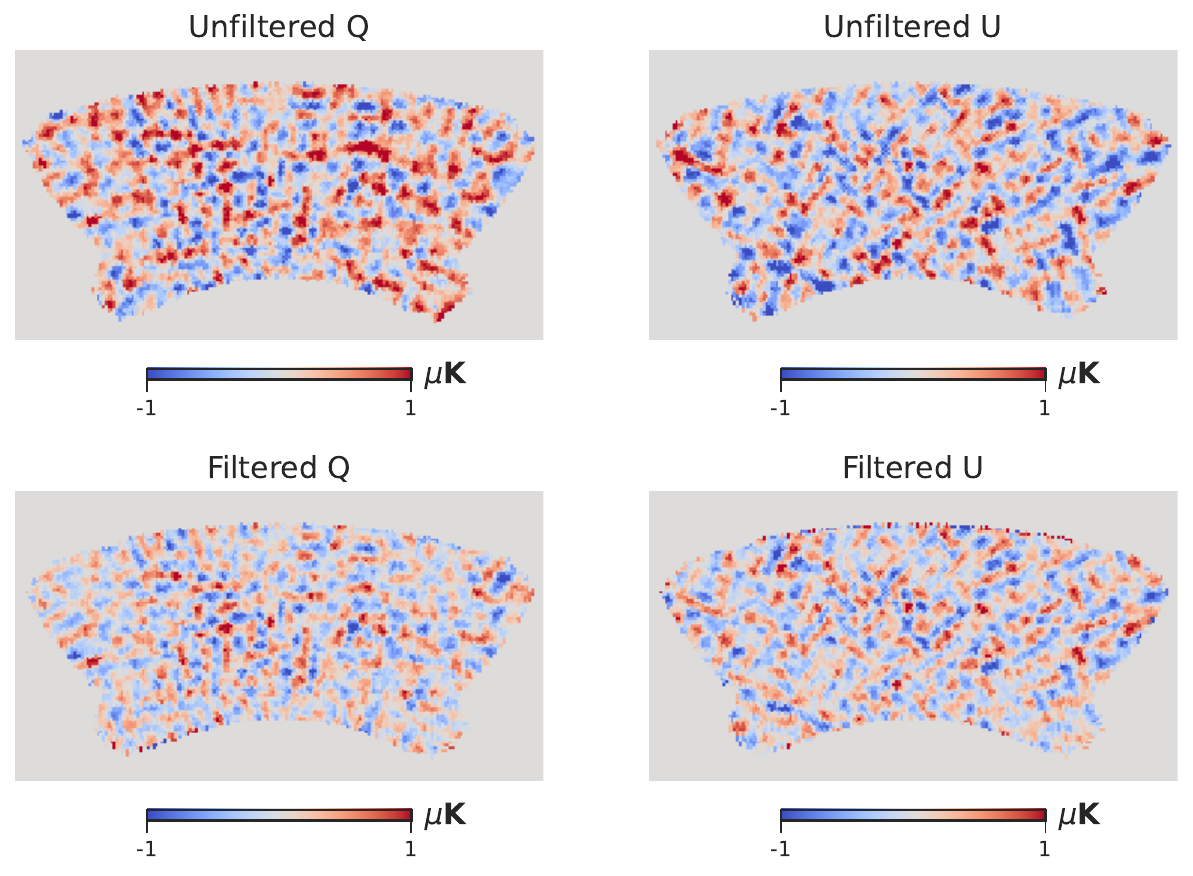}
\caption{Filtered vs unfiltered Stokes Q/U for a single signal-only EEBB realization. The filtering removes power on large scales, creating the ``washed out" effect seen in these maps.} 
\label{fig:filtered}
\end{figure}

The chosen filter stack is an order 1 polynomial subtraction for each subscan (CES between turnarounds) and an order 10 Legendre polynomial ground template (scan-synchronous signal) subtraction for the Q/U polarization TODs. These are standard filters that all ground-based experiments incorporate in some form and are known to cause E/B leakage due to their anisotropic nature. See Figure~\ref{fig:filtered} for a visual example of the filtering. 

\begin{figure}[htbp]
\centering
\includegraphics[width=0.6\linewidth]{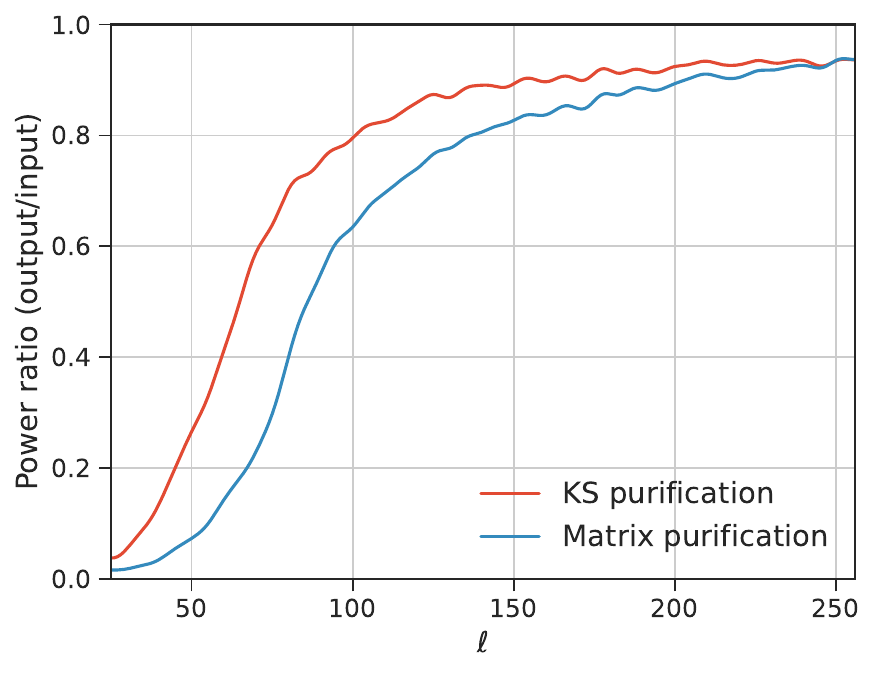}
\caption{The filter transfer function including KS and matrix purifications estimated from the mean of 1024 BB signal-only simulations. The filtering removes large scale power, reaching about 50\% signal attenuation around $\ell=60$ for KS purification. The matrix purification filter transfer function is lower in magnitude due to the removal of power from ambiguous modes.} 
\label{fig:ft}
\end{figure}

The BB filter transfer function $F_\ell$ is estimated iteratively with BB signal-only simulations, following \cite{Hivon_2002}
\begin{equation} \label{equ:ft}
    F_\ell^n = F_\ell^{n-1} + \frac{\Tilde{C}_\ell - \sum_{\ell'} \mathbf{M}_{\ell \ell'}F_\ell^{n-1}C_{\ell'}B_{\ell'}^2}{C_\ell B_\ell^2 f_\mathrm{sky}w_2}, F_{\ell}^0 = \mathbf{1}
\end{equation}
where $\Tilde{C_\ell}$ is the simulated pseudospectra, $\mathbf{M}_{\ell \ell'}$ is the mode mixing matrix, $B_\ell$ is the Gaussian beam function (with $\theta_\mathrm{fwhm}\approx1^\circ$), and $w_2$ is the second moment of the weight function. We find that 3 iterations gives stable results and use it to debias subsequent pseudospectra. Note that while not explicitly written in Equation~\ref{equ:ft}, we count the purification processes as part of the filtering, resulting in slightly different transfer functions for the KS and matrix purifications (Figure~\ref{fig:ft}).

\begin{figure}[htbp]
\centering
\includegraphics[width=0.9\linewidth]{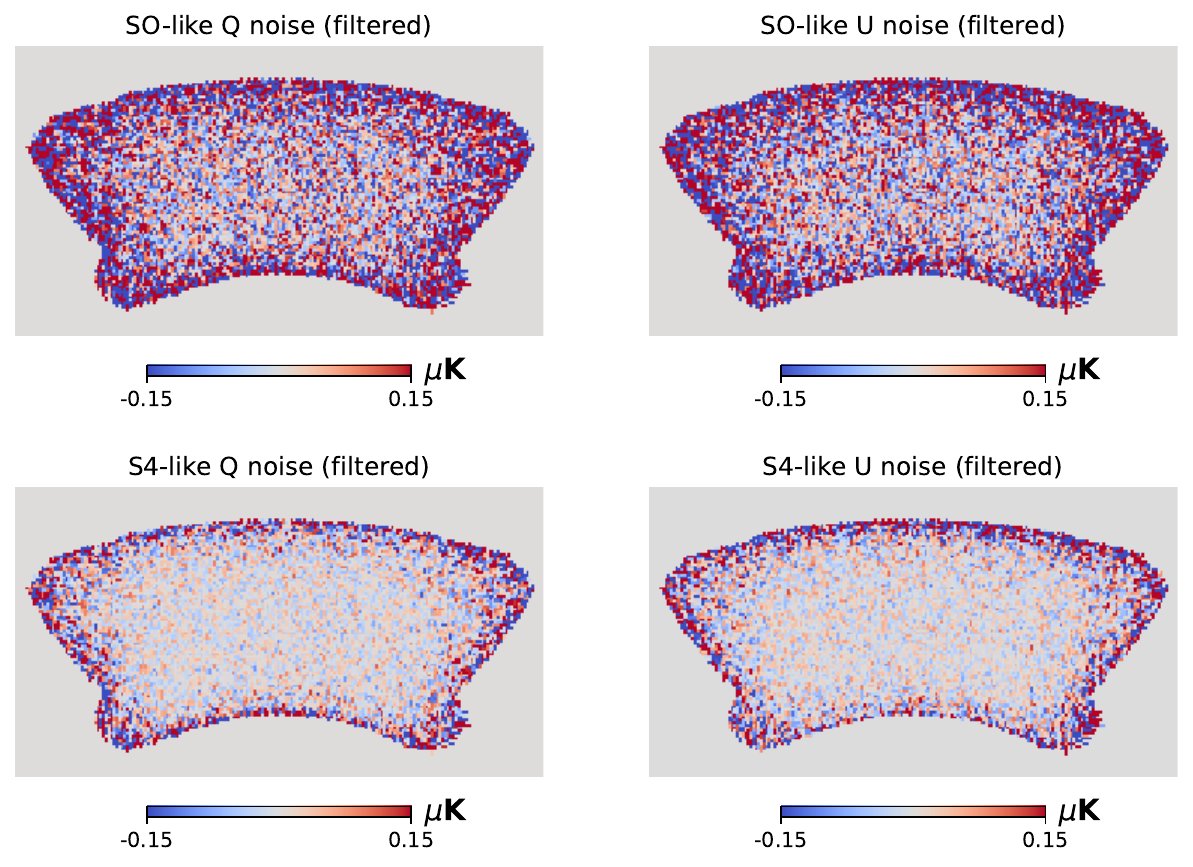}
\caption{Example simulated Q/U noise maps (filtered) of the SO-like and S4-like simulations, same color scale. The edges of the map are higher in noise due to the scanning coverage.} 
\label{fig:noise}
\end{figure}

The noise generation comprises an atmospheric noise simulation as well as a detector noise portion. The spectral properties (NET, $f_\mathrm{knee}$, $\alpha$) are adjusted to achieve the desired final map depth in polarization. We simulate two different map depths for the same patch in Figure~\ref{fig:hits} to approximate the noise spectra in Table~\ref{tab:depths}, allowing us to evaluate our E/B separation methods in the context of BB noise. The noise spectra $N_\ell$ follows the usual definition (without the beam) 
\begin{equation} \label{equ:nl}
    N_\ell = N_{w} + N_{w} \left( \frac{\ell}{\ell_\mathrm{knee}} \right) ^{\alpha}
\end{equation}
with $N_w$ being the white noise level set by the map depth. No correlated noise among detectors is simulated. See Figure~\ref{fig:noise} for an example noise realization. 

\begin{table}[htbp]
\centering
\begin{tabular}{c|c|c|c}
\hline
  & Map depth ($\mu$K-arcmin) & $\ell_{\mathrm{knee}}$ & $\alpha$\\
\hline
    SO-like pol & 3 & 50 & -3 \\
    S4-like pol & 1 & 50 & -3 \\
\hline
\end{tabular}
\caption{Two different simulated map depths in polarization for the ground-based experiment, one SO-like \cite{so} depth and one S4-like \cite{s4} depth. We note that in reality SO and S4 also differ in observed sky fraction, which is not taken into account here for ease of comparison.}
\label{tab:depths}
\end{table}

The final filter-binned ground-based experiment map for each Monte Carlo (MC) simulation is taken to be the sum of the observed signal-only map and the simulated noise map in Figure~\ref{fig:noise}.    

\subsection{Satellite E-mode Map Generation}
To generate noise for the satellite E-mode map, we first estimate effective noise spectra from experiments of interest and use \texttt{healpy.sphtfunc.synfast} to generate the noise maps themselves. As with the ground-based experiment, we simulate two different noise levels for the satellite E-modes (Table~\ref{tab:e-mode depth}).

\begin{table}[htbp]
\centering
\begin{tabular}{c|c|c|c}
\hline
 & Map depth ($\mu$K-arcmin) & $\ell_{\mathrm{knee}}$ & $\alpha$\\
\hline
    Planck-like temp & 25 & 50 & -1.8 \\
    Planck-like pol & 50 & 25 & -1.4 \\
    LiteBIRD-like temp & 1.8 & - & - \\
    LiteBIRD-like pol & 2.16 & - & - \\
\hline
\end{tabular}
\caption{Noise spectra parameters added to the satellite E-mode maps. Both temperature and polarization noise are simulated for the Wiener filtering step discussed below. The Planck-like values are estimated from the NPIPE2020 maps: \protect\url{https://portal.nersc.gov/project/cmb/planck2020/}. The LiteBIRD-like values \cite{litebird} assume white noise only due to the half-wave plate effectively eliminating the $\ell_\mathrm{knee}$. The map depth in temperature is simply scaled to be a $\sqrt{2}$ factor smaller than the polarization for simplicity --- it is inconsequential if LiteBIRD actually makes a temperature measurement.}
\label{tab:e-mode depth}
\end{table}

\begin{figure}[htbp]
\centering
\includegraphics[width=0.9\linewidth]{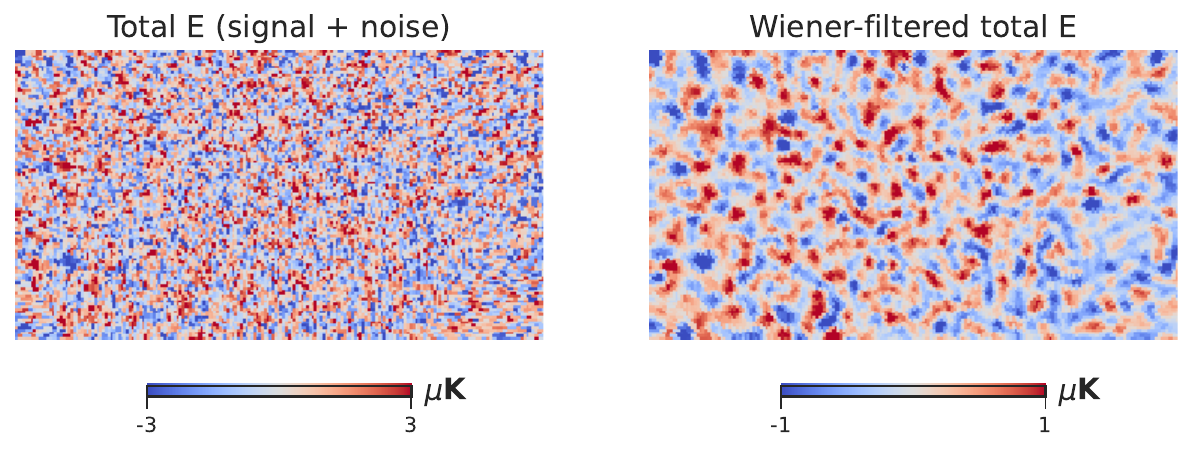}
\caption{A satellite E-mode map with Planck-like noise before (left) and after (right) Wiener filtering. Note the difference in the color scale. For illustration purposes, this view is plotted at the same zoom/resolution as we did with the ground-based experiment patch in previous figures.} 
\label{fig:wiener}
\end{figure}

\begin{figure}[htbp]
\centering
\includegraphics[width=0.6\linewidth]{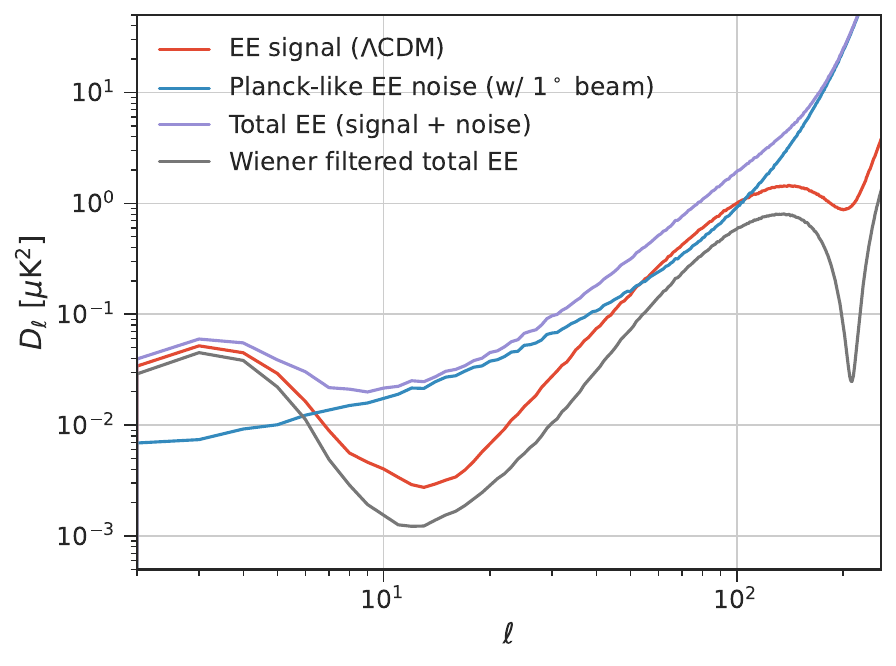}
\caption{EE power spectrum of 128 EE simulations + Planck-like noise for the satellite E-mode map, before and after Wiener filtering. The Wiener filter is effective in optimally removing noise while preserving signal.} 
\label{fig:PS_wiener}
\end{figure}

The noise map is added to the signal-only E-mode map to obtain the full satellite E-mode map with noise. Before we estimate the leakage template, we apply a multivariate Wiener filter to the noisy satellite E-mode map to obtain a minimum variance solution following \cite{Ghosh_2021}. This is especially important for the Planck-like E-modes, which have a signal-to-noise ratio of about unity. We have
\begin{equation}
    \mathbf{W} = \mathbf{S} (\mathbf{S} + \mathbf{N})^{-1} = 
    \begin{pmatrix}
        W_{\ell}^{TT} & W_{\ell}^{TE} \\
        W_{\ell}^{ET} & W_{\ell}^{EE}
    \end{pmatrix} 
\end{equation}
where \textbf{S} is the $2\times2$ TTEE beam-convolved input signal power spectra and \textbf{N} is the $2\times2$ TTEE noise power spectra that was simulated with the spectral properties in Table~\ref{tab:e-mode depth}. The Wiener-filtered E-mode spherical harmonic coefficients are then
\begin{equation}
    \widehat{d}_{\ell m}^E = W_{\ell}^{ET}d_{\ell m}^T + W_{\ell}^{EE}d_{\ell m}^E 
\end{equation}
where $d_{\ell m}^{T/E}$ are the spherical harmonic coefficients of the noisy map and $d_{\ell m}^B$ is set to zero. Including the temperature measurement recovers a significant amount of signal for the Planck-like case. $\widehat{d}_{\ell m}^E$ is then transformed back into Q/U space to use as our final satellite E-mode map. Figure~\ref{fig:wiener} and Figure~\ref{fig:PS_wiener} show the Wiener filter in action. 

\subsection{Implementation: Map-based E/B Separation} \label{subsec:map impl}
\begin{figure}[htbp]
\centering
\includegraphics[width=1\linewidth]{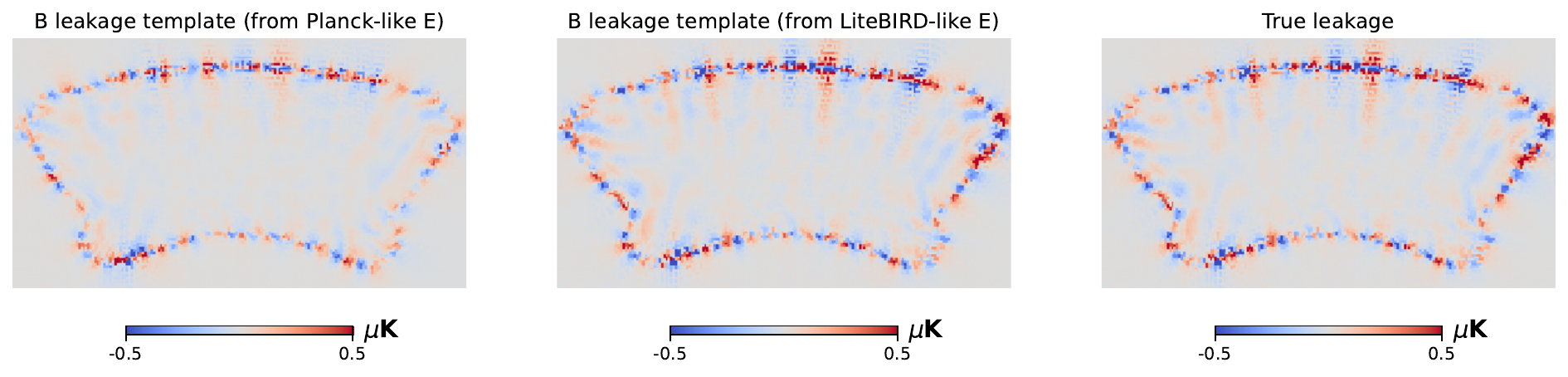}
\caption{B map leakage templates as a result of observing the Wiener-filtered E-mode maps. The leakage along the boundary comes from the partial sky coverage, whereas the dimmer structure inside the patch comes from the TOD filtering. The leakage template estimated from LiteBIRD-like E-modes (middle) is close to the true leakage (right), where the Planck-like E-modes (left) underestimate the leakage.} 
\label{fig:template}
\end{figure}

We now have all the tools to perform map-based E/B separation. The first step is to estimate the leakage template by processing our Wiener-filtered satellite E-mode map through the ground-based experiment TOD filtering pipeline. The resulting leakage template includes both the partial-sky and filtering leakages. See Figure~\ref{fig:template} for an example of the leakage templates estimated from Planck-like/LiteBIRD-like E-modes and the true leakage template. 

The estimated template is subtracted from the filtered ground-based experiment map in Q/U space to mitigate E/B mixing. We note that there are more nuanced approaches than map-space subtraction, such as template-fitting/cross-correlation methods in \cite{Ghosh_2022}, which we leave to future studies to compare. To be explicit, the exact operation to perform map-based leakage subtraction is
\begin{equation} \label{equ:subtract}
    \mathbf{m} = \mathbf{m}_\mathrm{obs} - \mathbf{m}_\mathrm{obs}^E   
\end{equation}
where $\mathbf{m}_\mathrm{obs}$ is the observed ground-based experiment Q/U map and $\mathbf{m}_\mathrm{obs}^E$ is the leakage template in Figure~\ref{fig:template} but in Q/U space. The leakage subtracted Q/U map is then passed to the power spectrum estimator, where KS purification is applied on top to tackle any residual leakage. We note that using KS purification in tandem with map-space leakage subtraction is not strictly necessary, and one can simply use the scalar-PCL method described in \cite{Ghosh_2021}. However, we have found that using both gives a slight edge in the results, which are discussed in Section~\ref{sec:results}.

\section{Results} \label{sec:results}
We explore the results in two parts. First, we evaluate the absolute value of the leakage for each mitigation method by looking at leakage B maps/BB spectra from EE simulations. Then, we compute the BB spectra from EEBB simulations plus ground-based experiment noise to put the results into context. For example, even if the absolute level of leakage between two methods differs by an order of magnitude, the impact on the final sensitivity can be negligible if both residual leakages are significantly below the noise level. We compute the Fisher uncertainty $\sigma(r)$ using the EEBB simulations as our final metric of merit. 

\subsection{Leakage Maps and Spectra (EE input)} \label{subsec:leakage}
\begin{figure}[htbp]
\centering
\includegraphics[width=1\linewidth]{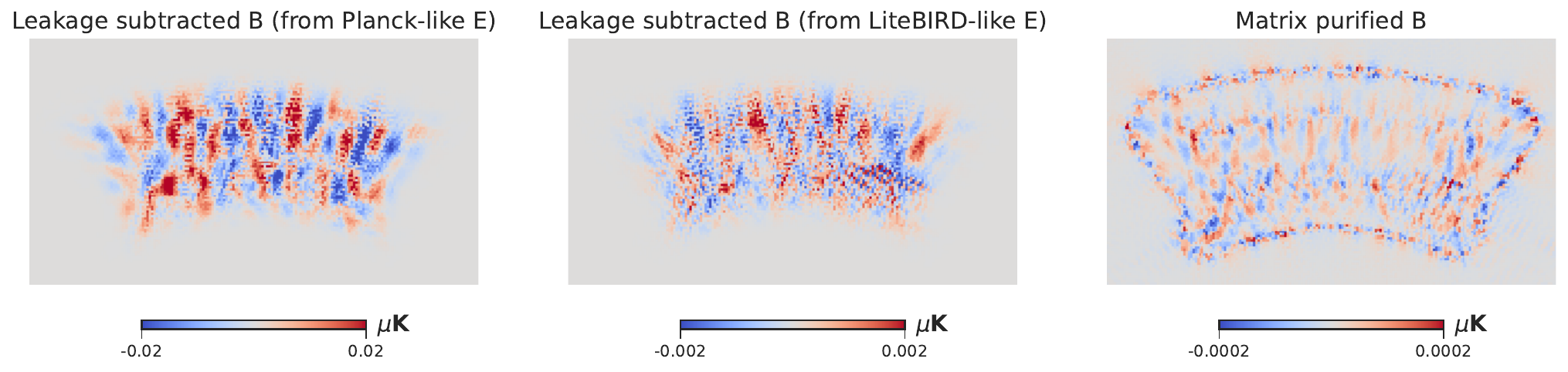}
\caption{Residual apodized leakage B maps after leakage subtraction with Planck-like E-modes (left), LiteBIRD-like E-modes (middle), and matrix purification (right). Note the dramatically different color scales. The matrix purification significantly outperforms both leakage template subtraction methods in the absence of noise and B-mode input.}
\label{fig:leakage_sub}
\end{figure}

\begin{figure}[htbp]
\centering
\includegraphics[width=0.6\linewidth]{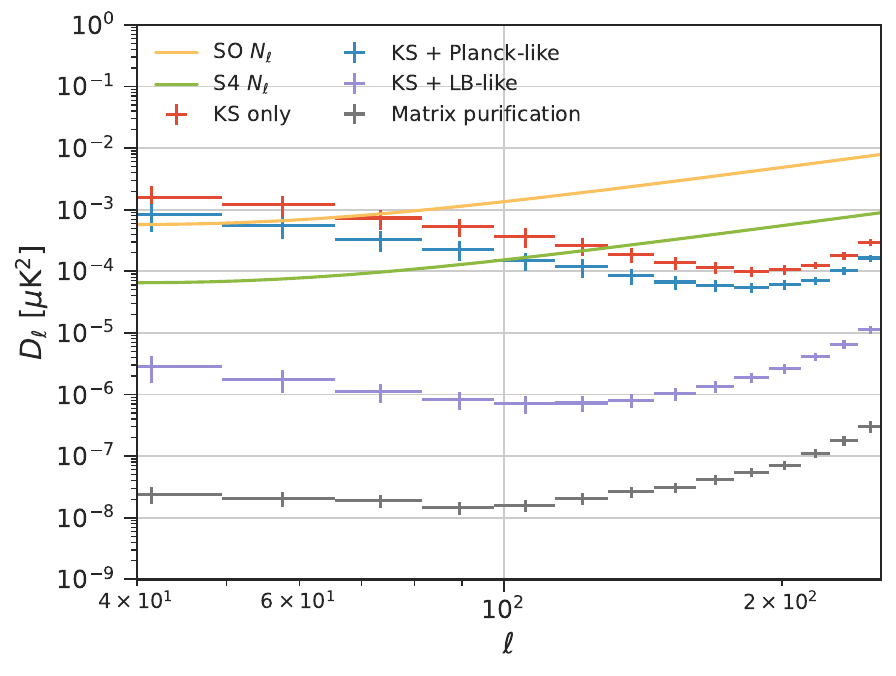}
\caption{BB leakage spectra for 256 EE simulations corresponding to the residual leakage maps in Figure~\ref{fig:leakage_sub}. Doing nothing to mitigate the TOD filtering leakage is labeled as ``KS only" (red) while leakage subtraction with the corresponding E-modes are labeled as ``KS + Planck/LB-like." As with the residual leakage maps, the matrix purification method performs the best in terms of absolute level of leakage. However, the residual leakage of both leakage subtraction with LiteBIRD-like E-modes and matrix purification are comfortably below the noise even for the S4-like case.}
\label{fig:PS_leakage}
\end{figure}

Our three E/B separation cases of study are:
\begin{enumerate}
    \item Residual B map/BB spectra of filtered EE input after leakage subtraction from Planck-like E-modes
    \item Residual B map/BB spectra of filtered EE input after leakage subtraction from LiteBIRD-like E-modes
    \item Residual B map/BB spectra of filtered EE input after matrix purification
\end{enumerate}
We examine their residual leakage maps/spectra in Figure~\ref{fig:leakage_sub} and Figure~\ref{fig:PS_leakage}, respectively.

We see from the color scales of the leakage maps in Figure~\ref{fig:leakage_sub} that both map-space leakage subtraction and matrix purification are effective in reducing leakage (compared to the unmitigated leakage in Figure~\ref{fig:template}). Template subtraction with LiteBIRD-like E-modes reduces the map-space leakage by an order of magnitude compared to Planck-like E-modes, and matrix purification performs the best of the three methods by another order of magnitude. This is corroborated by the leakage spectra results in Figure~\ref{fig:PS_leakage}. 

A practical factor that may degrade the performance of the matrix purification compared to our idealistic simulation is the accuracy of the observation matrix. Our simulations use a short 24-hour observing schedule for convenience, which means the observation matrix can be computed exactly. In practice, the observation matrix is estimated over a subset of the observing season with scanning representative of the entire dataset \cite{Ade_2016, bicep_thesis}, which may require more observing days from Chile compared to the South Pole. Another note is that due to the steep scaling of computational resources with $N_\mathrm{side}$, it is unlikely that the matrix purification can be done at the full resolution of the maps themselves. These factors are expected to have an impact visible on the scale of the leakage spectra in Figure~\ref{fig:PS_leakage}. However, as previously discussed, the absolute level of leakage is not always the most relevant metric in the presence of noise and other residuals. Since the residual leakage of both leakage subtraction with LiteBIRD-like E-modes and matrix purification are much lower than the noise levels of the ground-based experiment (Figure~\ref{fig:PS_leakage}), we expect both methods to perform similarly upon the inclusion of noise. In the next section, we will reevaluate the results in the presence of noise and lensing residuals.

\subsection{Spectra with Noise and Lensing Residuals (EEBB input)} \label{subsec:power spectra results}
We now consider input BB spectra with SO/S4-like noise levels combined with 10\% (nominal) and 1\% (optimistic) lensing residuals, which arise from delensing in the presence of noise \cite{s4}. We assume foreground residuals to be negligible and leave it for future work. 

\begin{figure*}[htbp]
\centering
\includegraphics[width=1\linewidth]{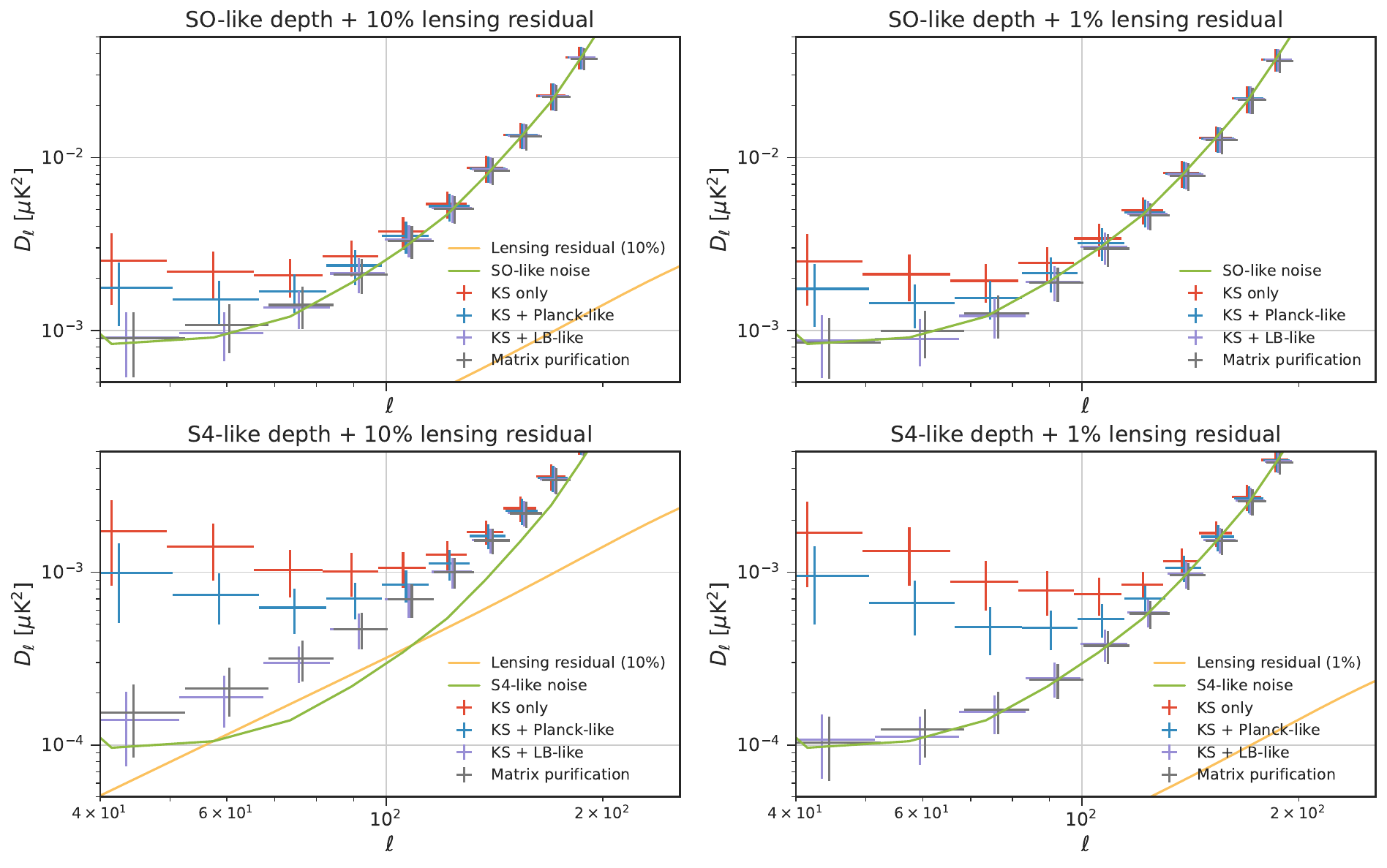}
\caption{Reconstructed BB spectra (256 EEBB simulations) of different E/B separation techniques in the context of two different map depths and lensing residuals. E/B leakage due to TOD filtering elevates the spectra above the noise bias starting around $\ell=100$. While the mean can be artificially debiased, the variance cannot, so we leave it in for illustration purposes. Each E/B separation technique can then be assessed on how close it brings the spectra back to the noise floor. A small $\Delta\ell=1$ offset is applied among the flavors in each bin to avoid overlap.}
\label{fig:ps}
\end{figure*}

Figure~\ref{fig:ps} is the culmination of our results. We see that the E/B leakage due to TOD filtering starts to bias the spectra at low-$\ell$, imparting significant additional variance. Leakage subtraction with Planck-like E-modes does only slightly better compared to KS purification only, which is expected given Planck's relatively high polarization noise. Leakage subtraction with LiteBIRD-like E-modes and matrix purification brings the spectra close to the noise floor once again. We see that despite the absolute level of leakage being orders of magnitude higher for the LiteBIRD-like case compared to the matrix purification, the difference ends up being negligible in the presence of noise and other residuals. To conclude our results, we propagate the spectra in Figure~\ref{fig:ps} to $\sigma(r=0)$.

\subsection{Fisher Forecast for $\sigma(r=0)$} \label{subsec:fisher}
\begin{figure*}[htbp]
\centering
\includegraphics[width=1\linewidth]{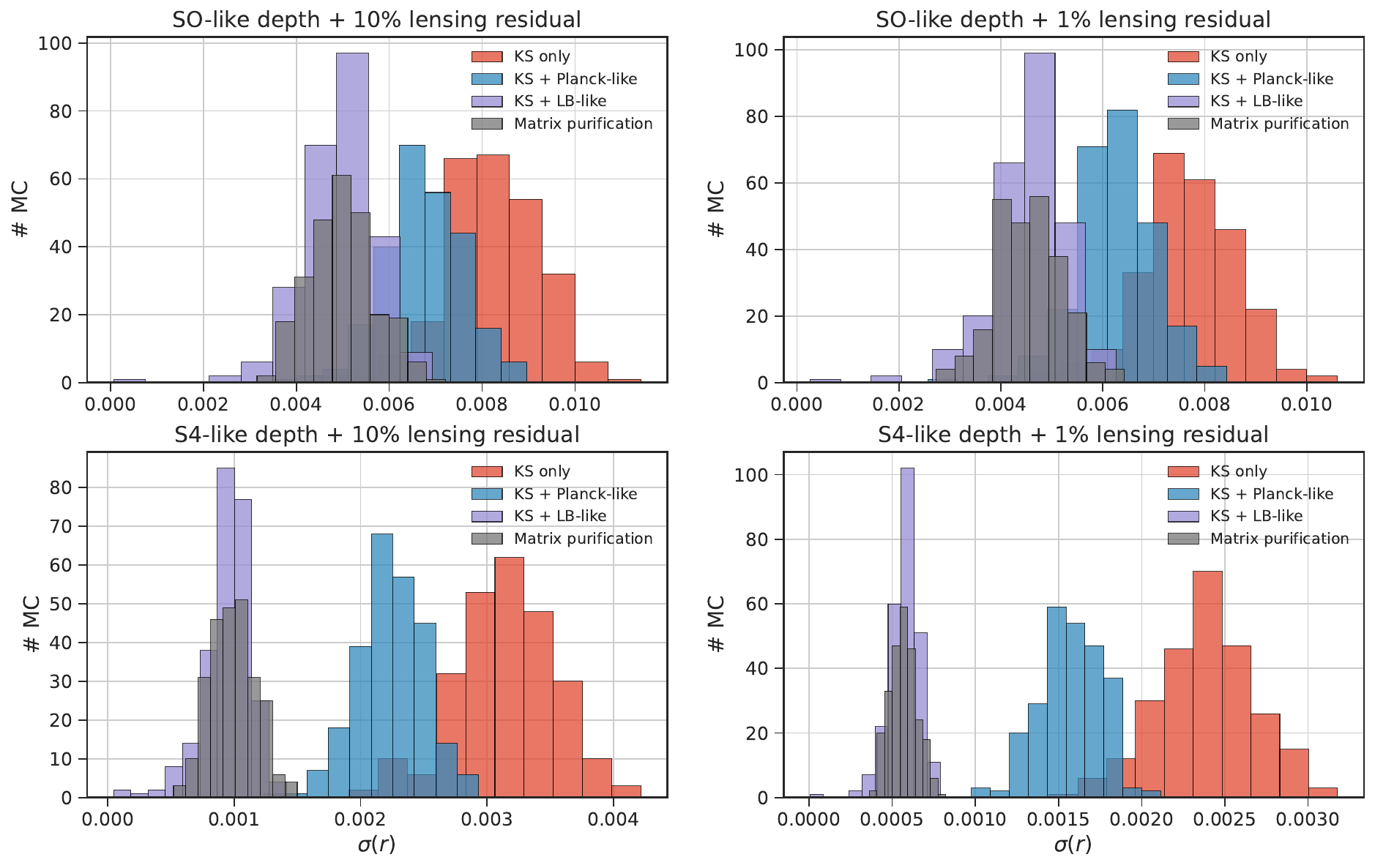}
\caption{$\sigma(r)$ histograms of each BB spectra in Figure~\ref{fig:ps}. Each E/B separation method improves upon $\sigma(r)$, shifting the histograms to the left. We note that the lower the noise/lensing residual, the bigger improvement each E/B separation method has. As already seen in Figure~\ref{fig:ps}, KS + LiteBIRD-like E-modes performs essentially just as well as the matrix purification for these residuals, at a much lower computational cost. The statistics are summarized in Table~\ref{tab:fish}.}
\label{fig:fish}
\end{figure*}

We first write the observed B-mode spectrum following \cite{Errard_2024} (ignoring foregrounds and other systematics)
\begin{equation}
    C_\ell^{BB,\mathrm{obs}} = r \times C_\ell^{BB,\mathrm{prim}}(r=1) + C_\ell^{BB,\mathrm{lens}} + C_\ell^{BB,\mathrm{noise}} + C_\ell^{BB,\mathrm{leakage}} 
\end{equation}
Given the likelihood $\mathcal{L}$ from \cite{Tegmark_1997}, the Fisher uncertainty $\sigma(r=0)$ can be written as
\begin{equation} \label{equ:fish}
\begin{aligned}
    \sigma(r=0) &\approx \left(\sqrt{\frac{\partial^2\mathcal{L}}{\partial r^2}\biggr\rvert_{r=0}}\right)^{-1} \\
    &\approx \frac{1}{\sqrt{\frac{f_\mathrm{sky}}{2}\sum_\ell(2\ell+1)\left(\frac{C_\ell^{BB,\mathrm{prim}}(r=1)}{C_\ell^{BB,\mathrm{lens}} + C_\ell^{BB,\mathrm{noise}} + C_\ell^{BB,\mathrm{leakage}}}\right)^{2}}}
\end{aligned}
\end{equation}
The BB spectrum from $40<\ell<256$ of each independent MC in Figure~\ref{fig:ps} is used as the denominator in Equation~\ref{equ:fish}, resulting in the histograms in Figure~\ref{fig:fish}. The percent improvement of $\sigma(r)$ given an E/B separation method $X$ over only using KS purification is defined as
\begin{equation} \label{equ:improv}
    \%\ \mathrm{improvement} = \frac{\lvert\sigma(r)^X_\mathrm{median} - \sigma(r)^\mathrm{KS}_\mathrm{median}\rvert}{\sigma(r)^\mathrm{KS}_\mathrm{median}} \times 100 
\end{equation}
and is shown for each method to the right of the arrow in Table~\ref{tab:fish} along with the summary statistics.

\begin{table}[htbp]
\centering
\begin{tabular}{c|c|c|c|c}
\hline
Method & SO-like/10\% & SO-like/1\% & S4-like/10\% & S4-like/1\% \\
\hline
    KS only & $8.2^{+1.0}_{-0.8}$ & $7.7^{+0.9}_{-0.8}$ & $3.1^{+0.4}_{-0.3}$ & $2.4^{+0.3}_{-0.3}$\\
    KS+Planck-like & $6.7^{+0.8}_{-0.8}\rightarrow18\%$ & $6.3^{+0.7}_{-0.7}\rightarrow18\%$ & $2.2^{+0.3}_{-0.2}\rightarrow29\%$ & $1.6^{+0.2}_{-0.2}\rightarrow33\%$\\
    KS+LiteBIRD-like & $5.0^{+0.6}_{-0.9}\rightarrow38\%$ & $4.6^{+0.6}_{-0.6}\rightarrow40\%$ & $0.96^{+0.2}_{-0.2}\rightarrow69\%$ & $0.58^{+0.07}_{-0.1}\rightarrow76\%$\\
    Matrix purification & $5.0^{+0.8}_{-0.7}\rightarrow39\%$ & $4.5^{+0.7}_{-0.6}\rightarrow41\%$ & $0.99^{+0.2}_{-0.2}\rightarrow68\%$ & $0.56^{+0.09}_{-0.08}\rightarrow76\%$ \\
\hline
\end{tabular}
\caption{$\sigma(r)\times10^3$ medians with 68\% percentile intervals of each histogram in Figure~\ref{fig:fish}. Rows correspond to E/B separation methods and columns correspond to the combination of a ground-based experiment noise level plus a lensing residual percentage. Improvement over KS only for a method $X$ is defined in Equation~\ref{equ:improv} and displayed to the right of the arrow in each cell. We note again that each E/B separation technique has an increasing factor of improvement with decreasing residual noise/lensing.}
\label{tab:fish}
\end{table}

Based on the percent improvement, we observe that cases with lower levels of residuals (such as S4-like depth + 1\% lensing) benefit more from all E/B separation methods and vice versa. This is the expected result when multiple residuals compete for critical path to optimal sensitivity, incentivizing us to carefully characterize our experiments.

\section{Discussion} \label{sec:discussion}
We have shown that the map-based E/B separation method via leakage template subtraction is effective in reducing $\sigma(r)$ by mitigating the E/B leakage due to TOD filtering. The absolute effectiveness depends on the depth of the E-modes used to estimate the leakage template, while the relative effectiveness compared to only using KS purification depends on the level of residuals present in the map.

In terms of absolute level of leakage mitigation, the matrix purification technique outperforms even leakage subtraction using LiteBIRD-like E-modes by two orders of magnitude (see discussion in Section~\ref{subsec:leakage}). However, since both levels of leakage are well below the noise levels of even the S4-like patch, they end up performing very similarly.

Considering the near future, experiments such as SO may be able to benefit from leakage subtraction using Planck E-mode maps by around 10-20\% in $\sigma(r)$ over using KS purification only, although we stress once again that our improvement factors in Table~\ref{tab:fish} are not meant to be a realistic forecast. Either way, we recommend that the matrix purification method be utilized if computational resources allow, as it vastly outperforms Planck E-modes in all cases. However, if there are additional systematics that increase the effective noise, we could be in a scenario in which we cannot reap the full benefits of the matrix purification. In that case, one would have to decide whether the computational challenges are worth the benefits.

Since the cost of implementing matrix purification will only increase as ground-based experiments deploy more detectors (see discussion of scaling in Section~\ref{subsec:matpure}), we imagine map-space leakage subtraction will become a much more attractive alternative once LiteBIRD (or similar depth) maps are available. The E/B separation performance is on par with matrix purification even at S4-like depths at a small fraction of the computational cost.

\subsection{Future Work}
We take the opportunity to discuss several avenues for improving the results in this paper: 
\begin{itemize}
    \item Use template-fitting/cross-correlation methods \cite{Ghosh_2022} instead of map-space leakage subtraction. This may help avoid the ``leakage" due to the satellite E-mode noise, which would be particularly important for the Planck E-modes.
    \item Use the Planck NPIPE2020\footnote{\url{https://portal.nersc.gov/project/cmb/planck2020/}} simulation maps to properly evaluate map-based E/B separation with Planck E-modes. Here, we simply generated isotropic noise for easier direct comparison with the LiteBIRD-like case. 
    \item Generate a more realistic observing schedule/patch geometry for forecasting. Use a subset of the observing schedule to compute the observation matrix and study the effect on matrix purification. We used a simple 24-hour schedule and a fixed patch for all experiments for simplicity, which allowed exact computation of the observation matrix. 
    \item Study the effect of instrumental systematics between ground-based and satellite experiments, such as calibration mismatch. Although ground-based experiments calibrate off of satellite experiments for their absolute gain, differences in relative gain will impact map-space combinations.
    \item Include foreground residuals as another source of $C_\ell^{BB}$, although this will involve the need to consider different observing frequencies. 
\end{itemize}

\section{Conclusion} \label{sec:conclusion}
E/B leakage due to TOD filtering is and will continue to be a critical source of sensitivity loss to mitigate for both current and future ground-based CMB experiments. However, the current standard of matrix purification may prove computationally challenging to scale into the future. We have presented an alternate method based on \cite{Ghosh_2021, Ghosh_2022} in which a leakage template is estimated from satellite E-modes and subtracted from ground-based experiment maps. The method was evaluated with a full TOD simulation for the ground-based experiment and compared to the existing matrix purification technique. We found that leakage subtraction with LiteBIRD-like E-modes and matrix purification perform very similarly for all cases of interest, with Planck-like E-modes lagging farther behind. As future experiments produce deeper E-mode maps, we expect to see map-based E/B separation become more mainstream due to its simplicity, high performance, and low computational cost.

\appendix
\section{Implementation: Matrix Purification} \label{sec:matpure impl}
We briefly describe our implementation of matrix purification used in this paper for completeness. We follow the procedure in \cite{Ade_2016, bicep_thesis} with slight adjustments. 

\begin{figure}[htbp]
\centering
\includegraphics[width=0.5\linewidth]{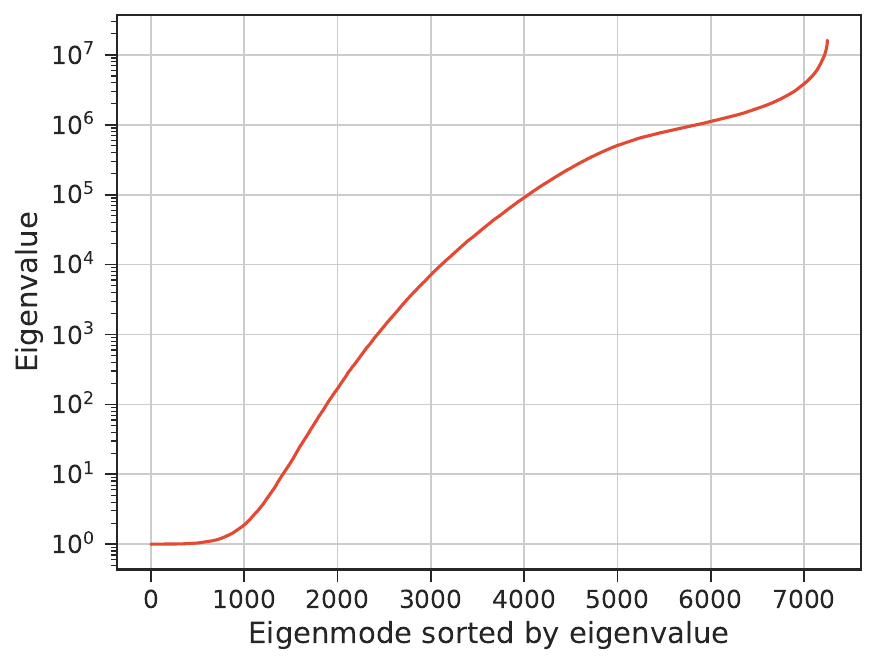}
\caption{Largest 7250 eigenvalues of the generalized eigenvalue problem (Equation~\ref{equ:gep}). Eigenmodes with larger eigenvalues are associated with more pure B-modes as per Equation~\ref{equ:eig_mag}.} 
\label{fig:eigs}
\end{figure}

\begin{figure}[htbp]
\centering
\includegraphics[width=0.9\linewidth]{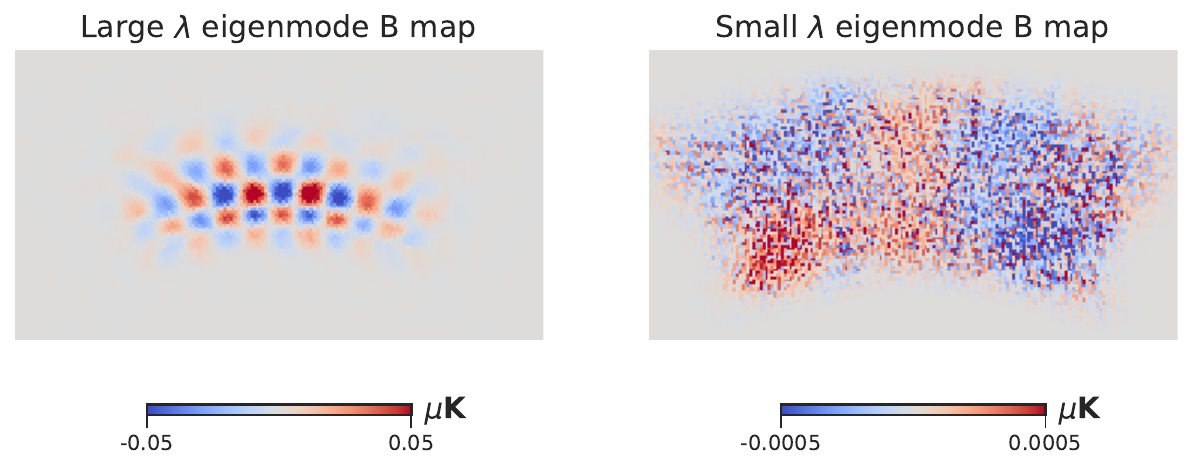}
\caption{Apodized eigenvectors with the largest eigenvalue (left) and the smallest eigenvalue (right) in our set, corresponding to the most ``pure" B-mode and the most ``ambiguous" mode in our set, respectively. We observe that the purer B-modes primarily manifest in low-$\ell$ power whereas more ambiguous modes contain more high-$\ell$ power.} 
\label{fig:eigv}
\end{figure}

The first step is to generate analytic signal covariance matrices $\mathbf{C}_E$ and $\mathbf{C}_B$ with a $1/\ell^2$ spectrum following \cite{Tegmark_2001}. We expedite the process by only considering the observed pixels in the hit map (Figure~\ref{fig:hits}). The observation matrix \textbf{R} with the selected observing schedule and filter stack is generated by TOAST and apodized with NaMaster
\begin{equation}
    \mathbf{R} = \mathbf{ZR}_{\mathrm{unapo}}
\end{equation}
where \textbf{Z} is the diagonal apodization matrix. The signal covariance is observed with \textbf{R} as per Equation~\ref{equ:obs_cov}. The generalized eigenvalue problem in Equation~\ref{equ:gep} is
\begin{equation} 
    (\Tilde{\mathbf{C}}_B + \sigma^2\mathbf{I})\mathbf{x}_i = \lambda_i (\Tilde{\mathbf{C}}_E + \sigma^2\mathbf{I}) \mathbf{x}_i
\end{equation}
where we perturb $\Tilde{\mathbf{C}}_{E/B}$ with a small regularization factor $\sigma$ (and \textbf{I} is the identity) to make the problem positive semi-definite. The factor $\sigma$ needs to be large enough so that the eigenvalue problem is numerically solvable, but not so large as to significantly perturb the generalized eigenvectors. We follow the BICEP/Keck collaboration's convention and set $\sigma$ to be 0.1\% of the mean of the diagonal of $\Tilde{\mathbf{C}}$. The upper half of the solved eigenvalue spectrum and the two extrema eigenvectors of our set are plotted in Figure~\ref{fig:eigs} and Figure~\ref{fig:eigv}, respectively.

Finally, the purification matrix is constructed via Equation~\ref{equ:pure}. We approximate the BICEP convention and use eigenvectors with $\lambda \geq 1.02$ to construct the purification matrix. The matrix-purified map is then just the product of the purification matrix with an observed map, as in Equation~\ref{equ:purify}.

\begin{figure}[htbp]
    \centering
    \includegraphics[width=1\linewidth]{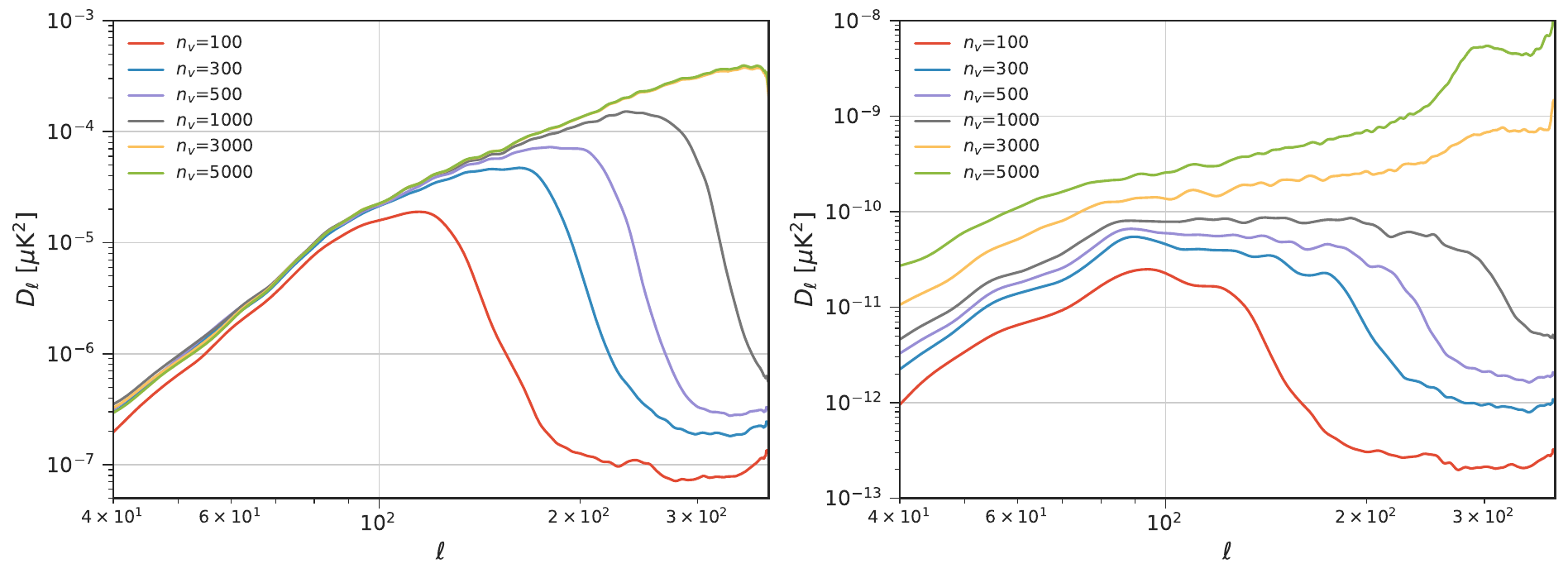}
    \caption{Left: Matrix-purified BB pseudospectra for 50 BB simulations with differing $n_v$. Higher $n_v$ includes more ambiguous modes, preserving more power at high-$\ell$ (see Figure \ref{fig:eigv}). Right: Matrix-purified BB ``leakage" pseudospectra for 50 EE simulations with differing $n_v$. Higher $n_v$ projects more power onto the ambiguous space, slightly increasing the level of E/B mixing. However, the absolute difference is usually negligible in the presence of the noise variance and other systematics.}
    \label{fig:mat_nv}
\end{figure}

We remark on the eigenvalue cutoff, or equivalently, the number of eigenvectors used to construct the purification matrix. We start at the eigenvector with the largest eigenvalue (the purest pure-B mode) and decide how many eigenvectors (in order of descending eigenvalue) to include in constructing the purification matrix. This number of eigenvectors, $n_v$, has an effect on the performance of the purification. Larger $n_v$ includes more eigenvectors with eigenvalues closer to unity (more ambiguous modes), which preserves more high-$\ell$ power at the cost of leakage mitigation (Figure~\ref{fig:mat_nv}).

\section{MC-based Matrix Purification}
\begin{figure}[htbp]
\centering
\includegraphics[width=1\linewidth]{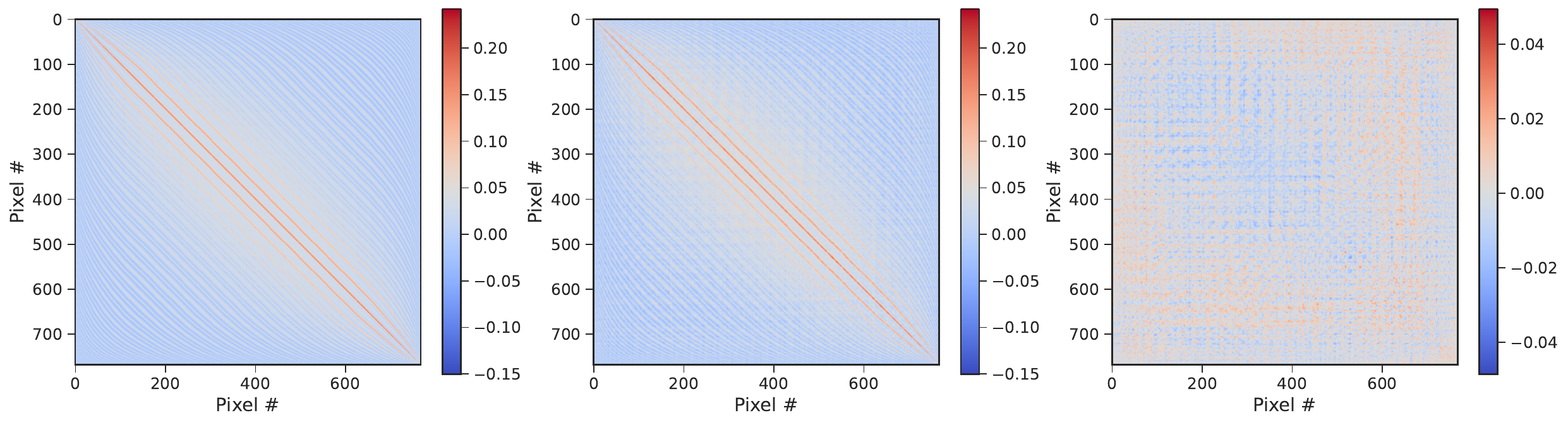}
\caption{QQ sector of analytic signal covariance matrix $\mathbf{C}_\mathrm{ana}$ (left), MC-based signal covariance matrix $\mathbf{C}_\mathrm{MC}$ (middle), and the difference map (right). The residual decreases with the number of simulations used to estimate $\mathbf{C}_\mathrm{MC}$.} 
\label{fig:cov_mc}
\end{figure}

Since there are many computational concerns with the construction of the observation matrix (see Section~\ref{subsec:matpure}), an alternative approach is to directly MC the observed covariance matrices in Equation~\ref{equ:obs_cov}. This involves generating filtered simulation maps and taking the mean of the outer product to construct the covariance matrix
\begin{equation}
    \mathbf{C}_\mathrm{MC} = \langle \mathbf{m} \mathbf{m}^\intercal \rangle
\end{equation}
where $\mathbf{m}$ is an observed signal only map. Although this method may also be computationally costly depending on how efficient the pipeline is, it is straightforward and easily parallelizable. The rest of the procedure is the same, except that the observation matrix and the signal covariance matrices are never explicitly constructed. Figure~\ref{fig:cov_mc} shows a visual example at $N_\mathrm{side}=8$ between the analytic and MC-based covariance matrices for the full-sky unfiltered case. 

As a rule of thumb, the MC-based covariance matrices should be estimated with on the order of number of observed pixels in the map. This ensures that $\mathbf{C}_\mathrm{MC}$ has the same matrix rank as $\mathbf{C}_\mathrm{ana}$ and percent level accuracy. 

\section{Low-$\ell$ Sensitivity}
\begin{figure}[htbp]
    \centering
    \includegraphics[width=0.5\linewidth]{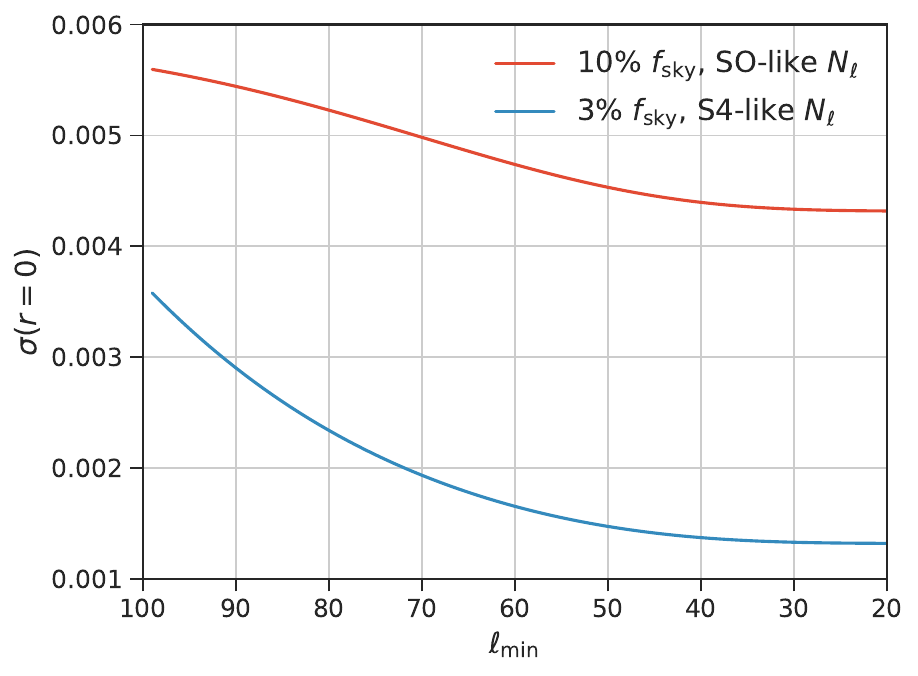}
    \caption{$\sigma(r)$ as a function of lowest $\ell$ in the power spectrum (decreasing). Improvement in $\ell_\mathrm{min}$ provides significant sensitivity gains above the $\ell=40$ range. Plotted for both SO and S4-like sensitivities (Table~\ref{tab:depths}).}
    \label{fig:lmin}
\end{figure}

To quantify low-$\ell$ sensitivity which E/B leakage due to TOD filtering negatively impacts, we plot Fisher uncertainty in Equation~\ref{equ:fish} as a function of the minimum $\ell$ considered in the power spectrum in Figure~\ref{fig:lmin}. For SO and S4-like sensitivities (Table~\ref{tab:depths}), improvement down to about $\ell_\mathrm{min}=40$ provides significant improvements in sensitivity.   

\section{Partial-sky (KS) Purification} \label{sec:ks_impl}
We demonstrate the built-in partial-sky purification functionality in the NaMaster software package \cite{Alonso_2019}, which we use in tandem with our map-based E/B separation method. 
\begin{figure}[htbp]
    \centering
    \includegraphics[width=0.5\linewidth]{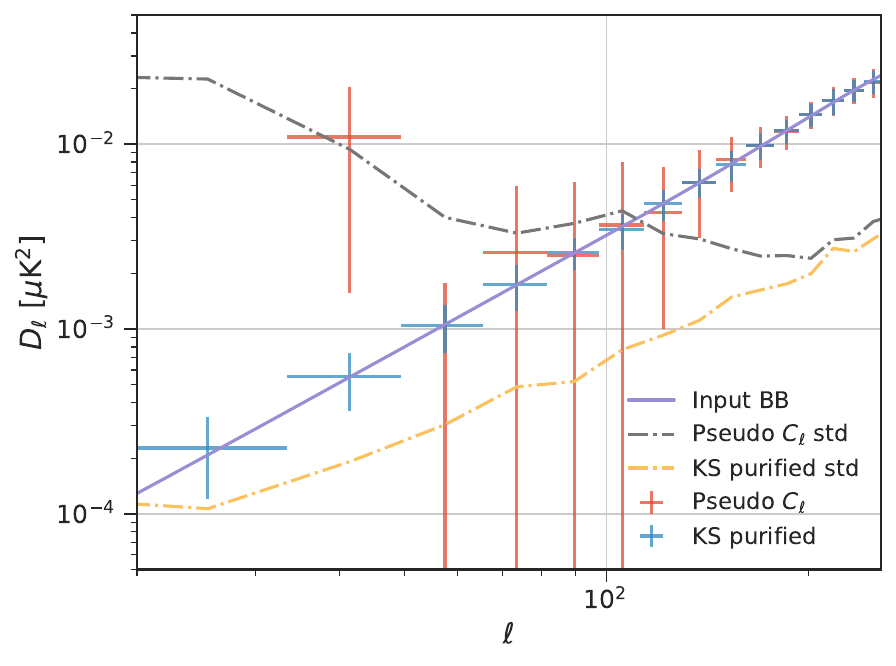}
    \caption{Comparison between KS purification (blue) and standard pseudo-$C_\ell$ (red) for the partial-sky case. Estimated from 128 masked EEBB simulations. The KS purified spectrum reduces the standard deviation by more than an order of magnitude at lower $\ell$ bins.} 
    \label{fig:KS}
\end{figure}

\acknowledgments
This work is supported by the Simons Array experiment funded by grants from the Simons Foundation, the Gordon and Betty Moore Foundation, the Templeton Foundation, the National Science Foundation, and the International Center for Quantum-field Measurement Systems for Studies of the Universe and Particles (QUP). Special thanks to Reijo Keskitalo of Lawrence Berkeley National Laboratory's Computational Cosmology Center for providing software support and observation matrix estimates. Lastly, we thank Shamik Ghosh for technical improvements, Josquin Errard for Fisher forecast discussions, Akito Kusaka for draft comments, and Radek Stompor for useful references. 



\bibliographystyle{JHEP}
\bibliography{biblio}
\end{document}